\documentclass[letterpaper,twocolumn,10pt]{article}
\usepackage[table,dvipsnames]{xcolor}
\usepackage{usenix2019_v3}
\usepackage[]{hyperref}
\hypersetup{
  colorlinks,
  linkcolor={blue},
  citecolor={red},
  urlcolor={blue!70!black}
}

\newenvironment{packeditemize}{
\begin{itemize}
  \setlength{\itemsep}{0.3pt}
  \setlength{\parskip}{2pt}
  \setlength{\parsep}{0pt}
}{\end{itemize}}

\usepackage{amsmath}
\usepackage{amssymb}
\usepackage{pifont}
\newcommand{\cmark}{\ding{51}}%
\newcommand{\xmark}{\ding{55}}%
\usepackage{makecell}

\usepackage{pgfplots}
\usetikzlibrary{shapes.geometric, arrows, external,calc}
\tikzstyle{kernel} = [rectangle, minimum width=3cm, minimum height=0.6cm,text centered, draw=black]
\tikzstyle{framework} = [rectangle, minimum width=3cm, minimum height=0.6cm,text centered, draw=black, fill=gray!30]
\tikzstyle{fs} = [rectangle, minimum width=3cm, minimum height=1cm,text centered, draw=black]
\tikzstyle{arrow} = [thick,->,>=stealth]

\definecolor{Red}{rgb}{0.8,0,0}
\newcommand{\todo}[1]{\textcolor{Red}{#1}}

\usepackage{pgfplots}
\pgfplotsset{compat=1.9}
\usepgfplotslibrary{external}
\usetikzlibrary{patterns}

\usepackage{soul}

    \tikzset{
        hatch distance/.store in=\hatchdistance,
        hatch distance=10pt,
        hatch thickness/.store in=\hatchthickness,
        hatch thickness=2pt
    }

    \makeatletter
    \pgfdeclarepatternformonly[\hatchdistance,\hatchthickness]{flexible hatch}
    {\pgfqpoint{0pt}{0pt}}
    {\pgfqpoint{\hatchdistance}{\hatchdistance}}
    {\pgfpoint{\hatchdistance-1pt}{\hatchdistance-1pt}}%
    {
        \pgfsetcolor{\tikz@pattern@color}
        \pgfsetlinewidth{\hatchthickness}
        \pgfpathmoveto{\pgfqpoint{0pt}{0pt}}
        \pgfpathlineto{\pgfqpoint{\hatchdistance}{\hatchdistance}}
        \pgfusepath{stroke}
    }
    \makeatother

\definecolor{cerise}{rgb}{0.87, 0.19, 0.39}
\definecolor{ceruleanblue}{rgb}{0.16, 0.32, 0.75}
\definecolor{ao(english)}{rgb}{0.0, 0.5, 0.0}
\definecolor{forestgreen(web)}{rgb}{0.13, 0.55, 0.13}
\definecolor{islamicgreen}{rgb}{0.0, 0.56, 0.0}
\definecolor{kellygreen}{rgb}{0.3, 0.73, 0.09}
\definecolor{limegreen}{rgb}{0.2, 0.8, 0.2}

\def\Snospace~{\S{}}

\usepackage[small, compact]{titlesec}
\usepackage{algpseudocode,amsmath}
\usepackage[noend]{algorithm2e}
\usepackage{titling}
\titlespacing*{\section}{0pt}{4pt}{3pt}
\titlespacing*{\subsection}{0pt}{3pt}{3pt}
\titlespacing*{\subsubsection}{0pt}{3pt}{3pt}

\usepackage[font=small]{caption}
\usepackage[font=small]{subcaption}
\usepackage{enumitem}
\setlist{nolistsep}

\usepackage{amsmath}
\usepackage[subtle]{savetrees}

\newcommand{\system}[0]{Bento}

\newcommand{\cut}[1]{}

\def\compactify{\itemsep=0pt \topsep=0pt \partopsep=0pt \parsep=0pt}
\let\latexusecounter=\usecounter

\titleformat{\section}
  {\normalfont\large\bfseries}{\thesection}{1em}{\MakeUppercase} 
\begin{document}

\date{}

\title{\Large \bf High Velocity Kernel File Systems with Bento\vspace{-5mm}}

\author{
 \normalfont
 Samantha Miller \enskip Kaiyuan Zhang \enskip Mengqi Chen \enskip Ryan Jennings \\ \normalfont Ang Chen$^\ddag$ \enskip Danyang Zhuo$^\dagger$ \enskip Thomas Anderson \\ \vspace{-2mm} \\
 \normalfont
University of Washington \enskip $^\dagger$Duke University \enskip $^\ddag$Rice University
\vspace{-10mm}
}

\pagenumbering{gobble}

\maketitle

\begin{abstract}

High development velocity is critical for modern systems. This is especially true for Linux file systems which are seeing increased pressure from new storage devices and new demands on storage systems.
However, high velocity Linux kernel development is challenging due to the ease of introducing bugs, the difficulty of testing and debugging, and the lack of support for redeployment without service disruption.
Existing approaches to high-velocity development of file systems for Linux have major downsides, such as the high performance penalty for FUSE file systems, slowing the deployment cycle for new file system functionality.

We propose \system{}, a framework for high velocity development of Linux kernel file systems. It enables file systems written in safe Rust to be installed in the Linux kernel, with errors largely sandboxed to the
file system. Bento file systems can be replaced with no disruption to running applications, allowing daily
or weekly upgrades in a cloud server setting. Bento also supports userspace debugging.
We implement a simple file system using \system{} and show that it performs similarly to VFS-native ext4 on a variety of benchmarks and outperforms a FUSE version by 7x on `git clone'.
We also show that we can dynamically add file provenance tracking to a running kernel file system 
with only 15ms of service interruption.
\end{abstract}

\section{Introduction}

Development and deployment velocity is a critical aspect of modern cloud software development. 
High velocity delivers new features to customers more quickly, reduces integration and debugging costs, and reacts quickly to security vulnerabilities.  However, this push for rapid development has not fully caught up to operating systems, despite this being a long-standing goal of OS research~\cite{chorus,mach,exokernel,spin,l4}. In Linux, the most widely used cloud operating system, release cycles are still measured in months and years. Elsewhere in the cloud, new features are deployed weekly or even daily.

Slow Linux development can be attributed to several factors. 
Linux has a large code base with relatively few guardrails, with complicated internal interfaces that are easily misused. Combined with the inherent difficulty of programming correct concurrent code in C, this means that new code is very likely to have bugs. The lack of isolation between kernel modules means that these errors often have non-intuitive effects and are difficult to track down. The difficulty of implementing kernel-level debuggers and kernel testing frameworks makes this worse.  
The restricted and different kernel programming environment also limits the number of trained developers.
Finally, upgrading a kernel module requires either rebooting the machine or restarting the relevant module, either way rendering the machine unavailable during the upgrade. In the cloud setting, this forces kernel upgrades to be batched to meet cloud-level availability goals.

Slow development cycles are a particular problem for file systems. Recent changes in storage hardware
(e.g., low latency SSDs and NVM, but also density-optimized QLC SSD and shingle disks) have made it
increasingly important to have an agile storage stack. Likewise, application workload diversity 
and system management requirements (e.g., the need for container-level SLAs, or  provenance tracking for security forensics) make feature velocity essential. Indeed, the failure of file systems to keep pace has led
to perennial calls to replace file systems with blob stores that would likely face many of the same challenges despite having a simplified interface~\cite{ceph}.

Existing alternatives for higher velocity file systems sacrifice either performance or generality. FUSE is a widely-used system for user-space file system development and deployment~\cite{fuse}.
However, FUSE can incur a significant performance overhead, particularly for metadata-heavy workloads~\cite{tofuse}. We show that the same file system
runs a factor of 7x slower on `git clone' via FUSE than as a native kernel file system. 
Another option is Linux's
extensibility architecture eBPF. eBPF is designed for small extensions, such as to implement a new
performance counter, where every
operation can be statically verified to complete in bounded time. Thus, it is a poor fit for implementing kernel modules like file systems with complex concurrency and data structure requirements. 


Our research hypothesis is that we can enable high-velocity development of kernel file systems without sacrificing performance or generality, for existing widely used kernels like Linux. Our trust model is that of a slightly harried kernel developer, rather than an untrusted application developer as with FUSE and eBPF. This means supporting a user-friendly development environment, safety both within the file system and across external interfaces, effective testing mechanisms, fast debugging, incremental live upgrade, high performance, and generality of file system designs.

To this end, we built Bento, a framework for high-velocity development of Linux kernel file systems. Bento hooks into Linux as a VFS file system, but allows file systems to be dynamically loaded and replaced without unmounting or affecting running applications except for a short performance lag.
As Bento runs in the kernel, it enables file systems to reuse well-developed Linux features, such as VFS caching, buffer management, and logging, as well as network communication.
File systems are written in Rust, a type-safe, performant, non-garbage collected language. Bento interposes thin layers around the Rust file system to provide safe interfaces for both calling into the file system and calling out to other kernel functions. Leveraging the existing Linux FUSE interface, a Bento file system can be compiled to run in userspace by changing a build flag. Thus, most testing and debugging can take place at user-level, with type safety limiting the frequency and scope of bugs when code is moved into the kernel. Because of this interface, porting to a new Linux version requires only changes to Bento and not the file system itself. Bento additionally supports networked file systems using the kernel TCP stack.
The code for Bento is available at \url{https://gitlab.cs.washington.edu/sm237/bento}.

We are using Bento for our own file system development, specifically to develop a basic, flexible 
file system in Rust that we call Bento-fs. Initially, we attempted to develop an equivalent 
file system in C for VFS to allow a direct
measurement of Bento overhead.  However, the debugging time for the VFS C version was prohibitive.
Instead, we quantitatively compare Bento-fs with VFS-native ext4 with data journaling, to determine 
if Bento adds overhead or restricts certain performance optimizations. 
We found no instances where Bento introduced overhead -- Bento-fs performed similarly to ext4 on most benchmarks we tested and never performs significantly worse while outperforming a FUSE version of Bento-fs by up to 90x on Filebench workloads. Bento-fs achieves this performance
without sacrificing safety. We use CrashMonkey~\cite{b3} to check the correctness and 
crash consistency of Bento-fs; it passes all depth two generated tests.
With Bento, our file system can be upgraded dynamically with only around 15ms of delay for running applications, as well as run at user-level for convenient debugging and testing. 
To demonstrate rapid feature development within Bento, we add file provenance tracking~\cite{linfs,pass} to Bento-fs and deploy it to a running system.

Bento’s design imposes some limitations. While Rust's compile-time analysis catches many common types
of bugs, it does not prevent deadlocks and or semantic guarantees such as correct journal usage---those errors must be debugged at runtime.
While correctness testing is possible at user-level, performance
testing generally must be done in the kernel. Also, like other live upgrade solutions, Bento upgrades also require backward-compatibility of the new code with
the previous data layout on disk---though the file system itself can perform disk layout changes.
The current implementation of Bento imposes some usability limitations similar to FUSE, such as only supporting
one mounted file system per inserted file system module.
And while we compare Bento-fs performance to ext4, we should note
that Bento-fs is a prototype and lacks some of ext4's more advanced features.


In this paper, we make the following contributions:

\begin{itemize}
    \item We design and implement Bento, a framework that enables high-velocity development of safe, performant file systems in Linux.
    \item We develop an API that enables kernel file systems written in a type-safe language with both user and kernel execution and live upgrade.
    \item We demonstrate Bento's benefits by implementing and evaluating a file system developed atop Bento with ext4-like performance, and show that we can add provenance support without rebooting. 
\end{itemize}

\section{Motivation}

Development velocity is becoming increasingly important for the Linux kernel to adapt to emerging use cases and address security vulnerabilities. In this section, we describe several approaches for extending Linux file systems, and outline the properties of \system{}.  

\subsection{High Velocity is Hard}

Linux needs to adapt to support emerging workloads, address newfound vulnerabilities, and manage new hardware. On average 650,000 lines of Linux code are added and 350,000 removed every release cycle, resulting in a growth of roughly 1.5 million lines of code per year. Linux file systems are no exception in needing
to adapt\,---\,with rapid change in both storage technologies and emerging application demands. 

As a concrete example, consider what is needed to add a feature like data provenance to a Linux file system. Increasingly, enterprise customers want to track the source data files used in producing each data analysis output file to perform security forensics. While this might be implemented with existing tools for system call tracking, that would be incomplete\,---\,the file system has more comprehensive information (e.g., whether two file paths are hard links to the same inode); a distributed file system can further enable cross-network forensics. To implement this as a new feature in the file system, developers have to modify the file system, test it, and push this modification to production clusters. 

The most widely used approach is to directly modify the kernel source code. 
Linux has standard kernel interfaces for extending its key subsystems\,---\,e.g., 
virtual file systems (VFS) for file systems, netfilter for networking, and Linux Security Module (LSM) for security features. Sometimes, it is also possible to add new features using loadable kernel modules, which can be integrated at runtime without kernel recompilation or reboot. Several VFS filesystems, including ext4, overlayfs, and btrfs, are implemented in the kernel source and can be inserted as loadable kernel modules.

However, high velocity kernel development (including kernel file system development) is hard to come by. 
To start with, kernel modifications are notoriously difficult to get right. Kernel code paths are complex and easy to accidentally misuse. Worse, debugging kernel source code is much harder than user-level debugging. This is because a kernel debugger operates below the kernel, typically remotely, and it cannot leverage Posix APIs such as ptrace. Upgrading kernel modules is also an intrusive operation. In the case of file systems, this requires shutting down applications, unmounting the old file system and remounting the new, and restarting the application. In a multi-tenant cloud setting, most cloud services
are upgraded live on a daily or weekly basis. To meet four or five nine application uptime
service-level objectives~\cite{nines} within a reboot model, however, kernel changes need to be batched 
and applied en masse every few months. Getting needed functionality upstreamed into Linux, so that
it is compatible with the 1.5M lines of new code being added each year, takes even longer.



\begin{table}[]
\small
\begin{tabular}{l l l}
\Xhline{4\arrayrulewidth}
\textbf{Bug} & \textbf{Number} & \textbf{Effect on Kernel}\\
\hline
Use Before Allocate & 6 & Likely \texttt{oops} \\
Double Free & 4 & Undefined \\
\texttt{NULL} Dereference & 5 & \texttt{oops} \\
Use After Free & 3 & Likely \texttt{oops} \\
Over Allocation & 1 & Overutilization \\
Out of Bounds & 4 & Likely \texttt{oops} \\
Dangling Pointer & 1 & Likely \texttt{oops} \\
Missing Free & 18 & Memory Leak \\
Reference Count Leak & 7 & Memory Leak \\
Other Memory & 1 & Variable \\
\hline
Deadlock & 5 & Deadlock \\
Race Condition & 5 & Variable \\
Other Concurrency & 1 & Variable \\
\hline
Unchecked Error Value & 5 & Variable \\
Other Type Error & 8 & Variable \\
\Xhline{4\arrayrulewidth}
\end{tabular}
\vspace{1mm}
\caption{Low-level bugs in released versions of OverlayFS, AppArmor, and Open vSwitch Datapath between 2014-2018,  categorized as memory bugs, concurrency bugs, or type errors, 
and the likely effect of each bug on kernel operation.\vspace{-2mm}} \label{tab:bugs}
\end{table}

To provide intuition into the difficulty of developing and deploying new kernel features, 
\autoref{tab:bugs} shows an analysis we conducted of bug-fix git commits from 2014-2018 for three 
modules that modify core Linux functionality used by Docker containers: 
OverlayFS, AppArmor, and Open vSwitch Datapath. We divide bugs in these systems
into two types. One set are semantic bugs in the high-level correctness properties of each module. 
These can range from mission critical to configuration errors, but generally impair just the functionality
of the module. These accounted for 50\% of the total bugs fixed in these modules.

The second set concern low-level bugs that are apply to any C language module, but when found
in the kernel can potentially undermine the correctness or operation
of the rest of the kernel. We categorized these as (1) memory bugs, such as \texttt{NULL} pointer dereferences, out-of-bounds errors, and memory leaks; (2) concurrency bugs, such as deadlocks and race conditions; and (3) type errors, such as incorrect usage of kernel types 
(e.g., interpreting error values as valid data). Of the 50\% of fixed bugs that were low-level bugs,
we found that 68\% are memory bugs. Of these, half are a type of memory leak. Many of the bugs occur in error-handling code, e.g., incorrect checking of return values, missing cleanup procedures. 
Such bugs are hard to uncover by testing but can lead to serious impacts on the integrity of the kernel. 
Of all identified low-level bugs, 26\% caused a kernel \texttt{oops} which either kills the offending process or panics the kernel. An additional 34\% of the analyzed bugs result in a memory leak, potentially causing out-of-memory problems or even DoS attack vectors. Many of these low-level bugs, particularly memory and type errors, result from inherent challenges of C code and could be prevented if the programming language had more safety checks.

\subsection{Existing Alternatives}

Besides directly modifying the Linux kernel, there are two other approaches to adding functionality
to Linux, with their respective pros and cons.

\paragraph{Upcall (FUSE~\cite{fuse}):}
One common technique, particularly for file systems and I/O devices, is to implement new functionality as a userspace server. A stub is left in the kernel that converts system calls to upcalls into the server. Filesystem in Userspace (FUSE) does this for file systems. As opposed to implementing new file system
functionality directly in the kernel, this isolates low-level memory errors such as use-after-free
to the userspace process. (Low-level bugs can still affect file system functionality, of course.) 
Development speed is faster because engineers can use familiar debugging tools like gdb. 
All this comes at a performance cost for metadata-operations~\cite{tofuse}. Our evaluation (\autoref{sec:microbenchmarks}) confirms this finding, revealing even worse performance overheads than previously reported, particularly for write-heavy workloads. Additionally, FUSE file systems can't reuse many existing kernel features, such as disk accesses through the buffer cache. Userspace file systems can mitigate the performance overhead by sharing mapped memory with the kernel, but this neither fully removes the performance overhead due to the extra kernel crossing nor allows the file system to access existing kernel functionality.

\begin{figure*}[t]
    \centering
    \includegraphics[trim=20 158 20 105, clip,width=0.75\textwidth]{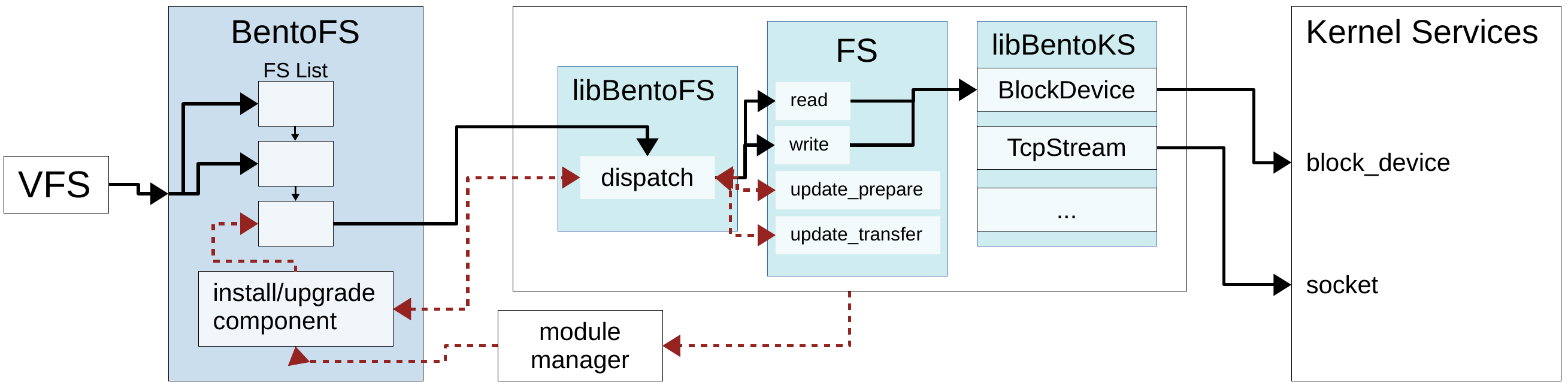}
    \caption{Design of Bento. Shaded components are parts of \system{}. \system{}FS is in C. The other shaded components are in Rust. Solid black lines represent the common-case operation pathway, detailed in \autoref{sec:fileops} and \autoref{sec:services}. Dashed red lines represent the install/upgrade pathway and are described in \autoref{sec:upgrade-component} \vspace{-2mm}}
    \label{fig:design}
\end{figure*}

\paragraph{In-Kernel Interpreter:}
Using an interpreter inside the kernel for a dynamically loaded program in a safe language is another approach to ensure safety of kernel extensions.
Linux supports eBPF (extended Berkeley Packed Filter)~\cite{bpf}, an in-kernel virtual machine that allows
code to be dynamically loaded and executed in the kernel at predefined points defined by the kernel.
eBPF is used heavily for packet filtering, system call filtering, and kernel tracing. 
The idea is to allow kernel customization in a safe manner. The Linux eBPF virtual machine validates
memory safety and execution termination before it JIT compiles the virtual machine instructions into native
machine code. As such, eBPF can sandbox untrusted extensions, but the restrictions placed
on eBPF make it very difficult to implement larger or more complex pieces of functionality.
We argue that untrusted eBPF extensions are not the right model for kernel file system extensibility,
as it is particularly difficult to imagine implementing mutable file 
system operations using eBPF and still enforcing crash consistency.

\subsection{Our Approach: Bento}  



We have designed \system{} for high velocity development of Linux file systems with 
the following properties: 

\begin{packeditemize}
\item \textbf{Safety:} Any bugs in a newly installed file system should be limited, as much as possible, to applications or containers that use that file system. 
\item \textbf{Performance:} Performance should be similar to that of the same functionality implemented using VFS.
\item  \textbf{Generality:} There are a large variety of file system designs that developers might want to implement. Bento should not limit the types of file systems that can be developed. 
\item \textbf{Compatibility:} File systems added to Linux via our framework should work with existing application binaries without recompiling or relinking. Further, Bento should not require 
substantial changes to Linux's internal architecture, to make Bento easier to upstream.
\item  \textbf{Live upgrades:} The framework should support dynamic upgrades to running file system code, transparently to applications, except for a small delay. 
\item  \textbf{User-level debugging:} File system code should be easily migrated between userspace and the kernel to enable user-level debugging and correctness testing. 
\end{packeditemize}

At a high level, \system{} achieves the first three goals by enabling developers to write file systems in Rust, a type-safe, non-garbage collected, general-purpose language that is receiving increasing attention for
kernel implementations. Of course, safely using Rust within Linux is a challenge of its own. The other three goals are achieved via careful architectural design. To provide \textit{compatibility} without sacrificing \textit{safety}, \system{} avoids directly using the Linux VFS interface, because it requires data structures to be directly passed back and forth between the file system and the kernel, making it difficult to provide verifiable data structure ownership safety. Instead, \system{} introduces a message-passing based API for file systems that enforces ownership safety. Second, \system{} introduces
a different API to enable safe access to C-language kernel services, by translating 
unsafe kernel interfaces into ones that can be safely used by Rust.
For \textit{live upgrades}, \system{} includes a component that quiesces the running file system and then transfers file system-defined state to the new instance, passing ownership of long-lived, in-memory data structures between the file systems so they can be shared across the upgrade.
For \textit{user-level debugging}, \system{} is designed with the same set of API calls whether it runs in the kernel or in the userspace. A simple build flag change is sufficient to choose a different mode.

\subsection{Rust Primer} \label{sec:rust}

As background, Rust is a strongly-typed, memory safe, data race free, non-garbage collected language. With these properties, Rust is able to provide strong safety guarantees without high performance overhead or the performance unpredictability caused by garbage collectors. These provide useful building blocks for \system{}. 

Rust relies on its type system to enforce memory safety.
The type system restricts how objects can be created and cast, so if an object exists and is of a certain type, this guarantees that the memory backing the object is valid and correctly represents that type.
Since raw pointers can be \texttt{NULL} and can be cast to nonequivalent types, dereferencing pointers and creating strongly type objects from pointers is unsafe and must be tagged as \texttt{unsafe} to compile. Calling unsafe functions is additionally unsafe. Although some systems allow unsafe Rust, Bento requires that its file systems contain no unsafe code.

Rust prevents most memory leaks by tracking the lifetime of objects.
All objects must be owned by one variable at a time. When the variable owning an object goes out of scope, the lifetime of the object is over and the memory backing the object can be safely reclaimed. References allow other variables to refer to data without claiming ownership of the memory. References are either immutable or mutable, enabling read-only or read-write accesses, respectively; references cannot outlive the owner. Developers can provide custom functionality to be performed when an object goes out of scope by implementing the \texttt{drop} method. Leaking memory is not a safety violation in Rust, so the \texttt{drop} function is not guaranteed to be called, but memory leaks must be explicit instead of accidental.

Data races are avoided by enforcing that all objects, except those that can be safely modified concurrently, must only have one mutable reference at a time.
For non-thread safe objects that must be shared between threads, synchronization mechanism such as locking must be used to safely obtain references. 
Acquiring the lock gives the caller access to the underlying data. 
Lock acquisitions methods generally return a guard that automatically unlocks the lock in \texttt{drop}, preventing the caller from forgetting to unlock.  However, deadlocks, such as circular waiting for locks,
are possible in safe Rust code as preventing them is beyond the power of the Rust type system. 
These represent about 7\% of the low-level bugs found in our analysis of popular kernel modules.

\section{The \system{} System} 

In this section, we describe the architecture of \system{}, explain how it interfaces with VFS and the rest of the kernel, and detail how it enables live upgrades and user-level debugging. 





\subsection{The System Architecture} 


\autoref{fig:design} shows the \system{} architecture; the shaded portions are the \system{} framework.
\system{} is a thin layer that, to the rest of Linux, operates like a normal VFS file system. 
The Linux kernel is unmodified other than the introduction of \system{}. In turn, like VFS, \system{} defines a set of function calls that \system{} file systems implement and provides a mechanism for file systems to register themselves with the framework by exposing the necessary function pointers. Unlike VFS, \system{} is designed to support file systems written in safe Rust.


\system{} consists of three components. First, \system{}FS interposes between VFS and the file system module and acts as a controller that manages registering and running file systems. \system{}FS is written in C and inserted as a separate kernel module. The other two components are Rust libraries that are compiled into the file system module. Lib\system{}FS translates unsafe calls from \system{}FS into the safe file operations API that is implemented by the file system. Lib\system{}KS provides a safe API for file systems to access kernel services, such as to perform I/O. The file system itself is written in safe Rust and is compiled as a Rust static library that includes lib\system{}FS and lib\system{}KS. When a file system module is loaded, it registers itself with \system{}FS which adds it to the list of active file systems.

\subsection{Interacting with VFS} \label{sec:fileops}

The VFS layer poses a fundamental challenge to memory safety. For example, 
VFS file systems allocate a single inode data structure to hold both VFS and file system-specific data.
When the kernel needs a new inode, it requests one from the file system which allocates it 
from its own memory pool. Both sides access their half of the data structure, and when done,
the kernel releases the inode to the file system so the memory can be reclaimed.
Independent of whether this is a good design pattern for minimizing kernel memory errors, it
is inconsistent with Rust compile time analysis and therefore would compromise our ability
to prevent memory safety errors within the file system code itself.

Instead, we define a new interface for safe kernel file systems. 
A selection of this API is in \autoref{tab:fileops-api}; the rest in the appendix.
The \system{}FS module receives all calls from the VFS layer, determines which mounted file system is the target, and handles any necessary operations on kernel data structures. \system{}FS then sends requests to the lib\system{}FS dispatch function using a similar API to that of the file system, but 
with unsafe pointers instead of Rust data structures. Lib\system{}FS parses the request, converts
pointers to safe data structures, and calls the correct function in the file system.
The key idea
is that the file system's compiler can statically verify its own data accesses, 
including its inode. To create an inode, \system{}FS calls into the file system (via lib\system{}FS)
and gets back an opaque reference (the inode number).
In turn, \system{}FS allocates and returns to VFS a separate kernel inode data structure. \system{}FS
never touches the contents of the file system inode.



\begin{table}[t]
    \small
    \centering
    \begin{tabular}{l}
        \Xhline{4\arrayrulewidth}
        \textbf{Bento File Operations API (partial)} \\
        \hline
        \textit{bento\_init(\&mut self, req, devname, fc\_info)} \\
        \textit{bento\_destroy(\&mut self, req)} \\
        \textit{bento\_read(\&self, req, ino, fh, offset, size, reply)} \\
        \textit{bento\_write(\&self, req, ino, fh, offset, data, flags, reply)}  \\
        \textit{bento\_update\_prepare(\&mut self) -> Option<TransferOut>} \\
        \textit{bento\_update\_transfer(\&mut, Option<TransferIn>)} \\
        \Xhline{4\arrayrulewidth}
    \end{tabular}
    \caption{A subset of the Bento File Operations API. 
    \textit{req} includes the user application's uid, gid, and pid. \textit{reply} includes data or error values. The full API is included in supplementary material.}  
    \label{tab:fileops-api}
\end{table}


\system{}FS and lib\system{}FS are responsible for ensuring that Rust’s safety properties are maintained as memory is passed across the File Operations API so the assumptions made by the Rust compiler will be true.
When passing references to kernel memory to the file system, such as data for read and write calls, \system{}FS guarantees that the memory will remain valid until the call completes and, if a mutable reference is passed, must ensure that no other thread is modifying the memory.
When passing references to structured data, \system{}FS and lib\system{}FS also ensure that the memory is correctly structured and never cast to an incompatible type.
Passing ownership across the File Operations API requires careful handling of the memory in lib\system{}FS and is only done during live upgrade (\autoref{sec:upgrade-component}).

\subsection{Interacting with Kernel Services} \label{sec:services} 

\system{} file systems need access to kernel functionality such as block I/O for access to underlying storage devices. These kernel interfaces, like those in the VFS layer, are not designed with type safety in mind and so cannot be directly used by a \system{} file system. Instead, lib\system{}KS implements safe versions of kernel data structures and functions needed by file systems. 

As an example, we will focus on kernel block I/O. File systems in Linux access block devices via the buffer cache. To read from (or write to) a block device, a Linux file system calls \texttt{\_\_bread\_gfp}, passing in a pointer to the \texttt{block\_device} data structure, a block number, the block size, and a page allocation flag. This function returns a \texttt{buffer\_head} data structure representing the requested block. The block's data is represented as a pointer and size in the \texttt{buffer\_head}. The file system can then read and/or write to this memory region. When the file system is done using the \texttt{buffer\_head}, it must call \texttt{brelse} or buffers can be leaked.


Like many kernel interfaces, kernel block I/O relies heavily on pointers. However, as described in \autoref{sec:rust}, raw pointers cannot be deferenced in safe Rust, and directly exposing these pointers to the file system results in safety errors. If the block I/O functions exposed to the file system accept a pointer, the block I/O functions cannot be marked safe and the file system as a whole cannot be safe.

\textbf{Exposing kernel services safely.} 
\system{} provides wrapping abstractions for kernel services so they can be used safely by the file system. These abstractions can be used like any other Rust data structures and functions. Several of the provided abstractions are detailed in \autoref{tab:kernel-services}.

\begin{table*}[]
\small
    \centering
    \begin{tabular}{p{18mm}l}
        \Xhline{4\arrayrulewidth}
        \textbf{Object Type} & \begin{tabular}{p{72mm}p{31mm}p{36mm}}
            \textbf{Method} & \textbf{Kernel Equivalent} & \textbf{Description} \\
        \end{tabular} \\
        \hline
        BlockDevice & \begin{tabular}{p{72mm}p{31mm}p{34mm}}
            \textit{bread(\&self, ...) -> Result<BufferHead>} & \textit{\_\_bread\_gfp(...)} & Read a block from disk \\
            \textit{getblk(\&self, ...) -> Result<BufferHead>} & \textit{\_\_getblk\_gfp(...)} & Get access to a block \\
            \textit{sync\_all(\&self) -> Result<i32>} & \textit{blkdev\_issue\_flush(...)} & Flush the block device \\
        \end{tabular} \\
        \hline
        BufferHead & \begin{tabular}{p{72mm}p{31mm}p{34mm}}
            \textit{data(\&self) -> \&[u8]} & \textit{buffer\_head->b\_data} & Get read access to data \\
            \textit{data\_mut(\&mut self) -> \&mut [u8]} & \textit{buffer\_head->b\_data} & Get write access to data \\
            \textit{drop(\&mut self)} & \textit{brelse(...)} & Release the buffer \\
            \textit{sync\_dirty\_buffer(\&mut self) -> Result<c\_int>} & \textit{sync\_dirty\_buffer(...)} & Sync a block \\
        \end{tabular} \\
        \hline
        GlobalAllocator & \begin{tabular}{p{72mm}p{31mm}p{34mm}}
            \textit{alloc(\&self, ...) -> *mut u8} & \textit{\_\_kmalloc(...)}/\textit{vmalloc(...)} & Allocate memory \\
            \textit{dealloc(\&self, ...)} & \textit{kfree(...)}/\textit{vfree(...)} & Free allocated memory \\
        \end{tabular} \\
        \hline
        RwLock<T> & \begin{tabular}{p{72mm}p{31mm}p{34mm}}
            \textit{new(data:T) -> RwLock<T>} & \textit{init\_rwsem(...)} & Create a RwLock of type \textit{T} \\
            \textit{read(\&self) -> LockResult<ReadGuard<'\_,T>{}>} & \textit{down\_read(...)} & Acquire the read lock \\
            \textit{write(\&self) -> LockResult<WriteGuard<'\_,T>{}>} & \textit{down\_write(...)} & Acquire the write lock \\
        \end{tabular} \\
        \hline
        TcpStream & \begin{tabular}{p{72mm}p{31mm}p{34mm}}
            \textit{connect(addr: SocketAddr) -> Result<TcpStream>} & $\left\{\begin{tabular}{@{}l}
            \textit{sock\_create\_kern(...)}  \\
            \textit{kernel\_connect(...)} \\
        \end{tabular}\right.$ & Create and connect \\
            \textit{read(\&mut self, ...) -> Result<usize>} & \textit{kernel\_recvmsg(...)} & Read a message \\
            \textit{write(\&mut self, ...) -> Result<usize>} & \textit{kernel\_sendmsg(...)} & Send a message \\
            \textit{drop(\&mut self)} & \textit{sock\_release(...)} & Cleanup the TcpStream \\
        \end{tabular}\\
        \hline
        TcpListener & \begin{tabular}{p{72mm}p{31mm}p{34mm}}
            \textit{bind(addr: SocketAddr) -> Result<TcpListener>} & $\left\{\begin{tabular}{@{}l}
            \textit{sock\_create\_kern(...)}  \\
            \textit{kernel\_bind(...)} \\
            \textit{kernel\_listen(...)}\ \\
        \end{tabular}\right.$ & Create, bind, and listen \\
            \textit{accept(\&self) -> Result<(TcpStream, SocketAddr)>} & \textit{kernel\_accept(...)} & Accept a connection \\
        \end{tabular}\\
        \Xhline{4\arrayrulewidth}
    \end{tabular}
    \caption{Kernel Services API. These are some of the data structures and methods provided to the file system. Methods that take \textit{\&mut self} can modify the object. Methods that take \textit{\&self} can access but not modify the object.}
    \label{tab:kernel-services}
\end{table*}

To be concrete, we address the example discussed above. We provide a safe \texttt{BlockDevice} abstraction to represent a kernel block device. A \texttt{BlockDevice} takes the name of the block device file and the block size; it contains a pointer to the kernel block device and the block size as fields. It provides several methods, including a safe \texttt{bread} method that takes a block number as an argument, performs safety checks, and calls \texttt{\_\_bread\_gfp} using the correct page allocation flag. The \texttt{bread} method returns a \texttt{BufferHead} that wraps the kernel \texttt{buffer\_head}. A \texttt{BufferHead} method converts the pointer and size fields into a sized memory region that can be used safely. That method must use unsafe code to make the sized memory region out of the unsized pointer and size fields, but the file system can call the method safely. To prevent accidental memory leaks, we call the \texttt{brelse} function in the \texttt{drop} method of the \texttt{BufferHead} wrapper. With this, buffer management has the same properties as memory management in Rust: memory leaks are possible but difficult.

Lib\system{}KS provides synchronization primitives including \texttt{RwLock<T>}, a wrapper around the kernel read-write semaphore. It has the same interface as the Rust standard library \texttt{RwLock<T>}, a read-write lock that protects data of type \texttt{T}. To obtain an immutable reference to the protected data, the user must acquire the read lock; to obtain a mutable reference, the user must acquire the write lock. \texttt{ReadGuard} calls \texttt{up\_read} in \texttt{drop} and \texttt{WriteGuard} calls \texttt{up\_write} in \texttt{drop}, preventing the user from forgetting to unlock.

In addition lib\system{}KS provides an implementation of the Rust global allocator that uses \texttt{kmalloc} and \texttt{kfree} for small regions (less than 8 pages) and uses \texttt{vmalloc} and \texttt{vfree} for larger regions. 
In this way, file system developers can use dynamically allocated types such as a growable array (Rust's \texttt{alloc::vec::Vec}) and collection types (from Rust's \texttt{alloc::collections}). Lib\system{}KS provides \texttt{TcpStream} and \texttt{TcpListener} to support networked file systems. 

These abstractions can, in some cases, add a small amount of performance overhead. If a kernel function has requirements on its arguments, the wrapping method likely will need to perform a runtime check to ensure that the requirements hold. 

\subsection{File System Upgrade} \label{sec:upgrade-component}

To enable online upgrades that are transparent to applications using the file system, we must first identify when it is safe to upgrade the file system and how to handle long-lived file system state. If an upgrade occurs while file system operations are still pending, there may be race conditions where some operations are executed on the old file system and others on the new, leading to correctness problems. In addition, any state that affects the semantic behavior of the file system, such as in-progress disk requests, file system journals, and TCP connections for networked file systems, must be correctly preserved across the upgrade. State that affects performance but not semantics, such as clean data in caches, can be optionally preserved.




\system{} addresses these challenges by ensuring that the old file system is in a quiescent state and that semantic state is transferred to the new file system. \system{} quiesces the file system by pausing new calls into the file system module during the upgrade and waiting for in progress operations to complete.
To achieve this, \system{} uses a read-write lock on the file system connection. All calls into lib\system{}FS acquire the read lock, while upgrades acquire the write lock. Therefore, file system operations can be executed concurrently in normal mode but will be blocked during an upgrade;
the upgrade will be blocked until previous operations complete.


Second, a constraint on the old file system is that it must be able to transfer its semantic state
to the new file system. Of course, the specific content of this state will vary 
from file system to file system. 
Each file system defines two data structures: one that is returned when the file system is removed and one that is expected when the file system is replacing a previous live file system. This design pattern,
of needing to write code to support both past and future versions, is common in cloud settings. During upgrade, ownership of the data structure is passed from the old file system to the new one. \system{}FS handles passing the data structure from the old file system to the new file system.
The detailed mechanisms involved for live upgrades are shown in \autoref{fig:design} and described below: 
\begin{enumerate}
    \itemsep-0.05em
    \item A new file system upgrade instance is loaded into the kernel. At module load, it calls into \system{}FS to register itself and indicate that it is an upgrade.
    \item \system{}FS identifies the file system that needs to be unloaded and acquires the lock to
    pause new operations and wait for existing operations to complete.
    \item \system{}FS sends a \texttt{bento\_update\_prepare} request to the old file system through lib\system{}FS.
    \item The old file system instance handles the \texttt{bento\_update\_prepare} request, performing any necessary cleanup and creating and returning its defined output state transfer struct to \system{}FS through lib\system{}FS.
    \item \system{}FS sends a \texttt{bento\_update\_transfer} request to the new file system through lib\system{}FS, passing the state transfer data structure to the new file system.
    \item The new file system instance initializes itself using the provided state and returns.
    \item \system{}FS modifies the connection state by replacing the old file system reference with the new file system reference and releases the write lock, allowing calls to proceed to the new instance.
\end{enumerate}




\subsection{Userspace Debugging Support} \label{sec:userspace-debugging}



\system{} also introduces a feature that enables a new file system to be seamlessly hoisted to userspace for debugging. This enables developers to leverage gdb and other familiar utilities for higher velocity development. Debugged code can then be dropped back into the kernel without any modification. \system{} supports this feature by exposing identical interfaces to both the kernel version and the userspace version of a developed file system. Whether the file system runs in the kernel or at userspace is determined by a compilation configuration flag which specifies which libraries will be linked and how the file system should register itself during initialization. 

Our solution leverages Linux kernel FUSE support to forward file operations to userspace. 
By itself, this is not sufficient\,---\,a FUSE file system is not runnable in the kernel.
At a high level, we design our kernel interfaces to mirror existing userspace interfaces when possible, and implement userspace libraries to expose additional abstractions otherwise. 

Many kernel interfaces can be designed to expose the same interfaces as userspace abstractions. For example, kernel read-write semaphores are used the same way as Rust’s \texttt{std::sync::RwLock<T>} and the kernel TCP stack provides similar interfaces to Rust’s \texttt{std::net::TcpStream} and \texttt{std::net::TcpListener}. In these cases, our kernel services API provides interfaces that are identical to the analogous userspace interface.

However, some kernel interfaces do not have obvious userspace analogues. The File Operations API (\autoref{tab:fileops-api}), for example, 
adds functions to implement state transfer and passes immutable references to ensure correct concurrency behavior. Additionally, operations on the backing storage device are performed differently from the kernel and userspace. FUSE file systems typically use file I/O to access the storage device while kernel file systems directly interface with the kernel buffer cache. Using a file I/O interface in the kernel would significantly hinder performance and functionality, adding extra data copies and preventing certain optimizations. However, there is no standard userspace abstraction that closely mirrors the kernel buffer cache.

To address this, we provide two additional libraries 
The userspace version of lib\system{}FS translates calls from FUSE into the File Operations API.
The userspace version of lib\system{}KS implements a basic buffer cache that uses file I/O under the hood, providing the \texttt{BlockDevice} and \texttt{BufferHead} abstractions to \system{} file systems
when running at user level.

\section{Implementation \& Experiences} 

We have developed \system{} as a Linux kernel module for \system{}FS and a Rust library containing both lib\system{}KS and lib\system{}FS in 5240 lines of C and 5072 lines of Rust. The userspace versions of lib\system{}KS and lib\system{}FS are another 986 lines of Rust. The current implementation targets Linux kernel version 4.15. The file system is compiled as a Rust a static library, which can be linked with any required C code to generate the .ko kernel module. Kernel code in Rust cannot use standard libraries, but we do enable use of the Rust alloc crate.


 \subsection{\system{}FS}
 We built \system{}FS by modifying the existing Linux FUSE kernel module. 
 In place of upcalls, \system{}FS communicates with lib\system{}FS using function calls. A file system module registers itself with \system{}FS by providing a pointer to the \texttt{dispatch} function when it is mounted. Like the VFS layer, \system{}FS maintains a list of active file systems, locking the list and adding and removing entries when file systems are registered or unregistered. This list is additionally locked during a live upgrade.
 
 \textbf{Upgrade State Transfer. }
Ownership of state transfer data structures must be moved between the Rust file system modules during an upgrade to allow the new file system instance to take ownership of state owned by the old file system instance. We implement this ownership transfer in lib\system{}FS using the Rust \textit{Box} type. 
When the old file system instance returns its state to lib\system{}FS, we create 
a \textit{Box} to take ownership of the data and pass the box as a raw pointer to \system{}FS.
The new lib\system{}FS converts the pointer back to a \textit{Box}, 
claiming ownership of the data before passing it to the file system. 
Rust deletes the old file system data structure when it goes out of scope at the end of the transfer;
the old file system is uninstalled in the background.
\subsection{Experiences Using Bento}

We began this project developing both a Bento version of a file system and its VFS equivalent in C, as
a way to quantify the performance cost of Bento. However, we eventually stopped development on the VFS version because implementing and debugging new features were significantly more time consuming and difficult than for the Bento version. In VFS, we were much more likely to accidentally write memory errors, such as \texttt{NULL} pointer dereferences and memory leaks. These bugs took much longer to diagnose and fix than bugs in the Bento version because they would crash the kernel, forcing us to reboot between tests, and they were difficult to isolate.

We further illustrate our experience developing with \system{} on three axes: functionality, performance, and correctness.

\textbf{Functionality. }
Using \system{}, we implemented \system{}-fs, a file system designed to have ext4-like performance, in 3038 lines of safe Rust code. \system{}-fs is structurally similar to the xv6 file system, a simple file system included in MIT's teaching operating system xv6~\cite{xv6}. This simplicity made the xv6 file system an attractive starting point for our prototype. \system{}-fs includes several modifications for improved functionality and performance. For example, xv6 does not fully support the functionality necessary to run our benchmarks. Likewise, we added double indirect blocks to 
support files up to 4GB, instead of 4MB in xv6.

We also added a provenance feature to \system{}-fs. The architecture of provenance tracking is borrowed from existing work~\cite{linfs, pass}. It consists of two pieces: a) a file system component that tracks file creations, deletions, and opens; and b) a syscall-level component that tracks the process hierarchy and operations on open file descriptors, such as dup and sendmsg. 

The file system-level component is implemented by logging information to a special file. To track existing files, `create', `rename', `symlink', and `unlink' operations log the user process ID of the request, the names and inode numbers of relevant files, any request flags, and, for `unlink', whether or not the file was deleted. The current implementation does not track hard links, but adding such support could follow a similar strategy. Since Bento-fs is not called for every read or write operation due to kernel caching, we track file accesses by logging `open' and `close' calls, recording the read/write mode of the open call along with the process ID of the request and the inode number of the file. If a file is opened as writable while another file is opened as readable, provenance tracking assumes that the writable file’s contents depends on the readable file’s contents.

The syscall-level component tracks process creation through `fork'/`exec' and operations on open file descriptors so the provenance system can correctly handle instances where a process gains access to a file without using the open syscall. This component is implemented as a collection of eBPF programs that log the relevant system calls, namely `clone', `exec', `pipe', `dup', `dup2', and `sendmsg'. `Open' calls are also logged so the file descriptors used in the system calls can be matched to the file system tracking on file names.

Overall, these features were added to Bento-fs in 145 lines of code in two weeks of development. In our development process, we never caused a crash of the operating system and were able to test and debug code within minutes of making changes. In fact, many of our changes worked correctly once they compiled, something that has not been true of our C development.

\textbf{Performance. }
To be able to bound the overhead imposed by Bento by comparing it to ext4, 
we added various optimizations to Bento-fs to match ext4 behavior.
We particularly noticed overhead on multi-threaded and metadata intensive benchmarks.
The xv6 free inode and free block implementations, for example, are needlessly inefficient.
The journal used by xv6 is small by default and assumes that each operation will use the maximum number of blocks, limiting it to only three concurrent operations at once. It also commits operations to the device synchronously when transactions are completed. 
We increased the size of the log and leveraged the Linux journal module JBD2 (also used by ext4).
In JBD2, transactions request the required number of blocks and commit in the background. \footnote{Although we implemented a log manager for the userspace version, it is likely less optimized than the kernel version, and there may be additional ways to improve userspace write performance that we have not yet discovered.}

Similarly, xv6 uses an inefficient list structure for directories. We added tree-structured directories that use the hash of the file name to locate directory entries. 

Most of the code changes for the journal modifications were in lib\system{}KS and mkfs. Tree structured directories were implemented within \system{}-fs in around 800 lines of code, split across utility functions for the hash tree and directory lookup, linking, and reading. Having access to dynamically allocated data structures from Rust's \texttt{alloc} crate simplified this implementation.
The tree structure uses the B-tree implementation provided by the crate and the directory lookup, linking, and reading code use Rust's dynamically allocated array \texttt{Vec}.

\textbf{Correctness. }
We tested the correctness of our file system using CrashMonkey~\cite{b3}. It generates workloads based on operations supported by the file system, and exhaustively tests all combinations up to a defined sequence length. We ran the \texttt{seq-2} benchmarks~\cite{b3}, which test sequences of two operations, using the operations supported by \system{}-fs. This resulted in 47314 benchmarks in total.
CrashMonkey did not find any crash consistency bugs in \system{}-fs. It found a known bug from the FUSE kernel module in the C code used in \system{}FS where opening a directory then calling \texttt{rmdir} followed by \texttt{mkdir} on the directory name before closing it resulted in an unusable directory due to inode reuse. We fixed this by always allocating a new inode during directory creation.

The provenance extension to Bento-fs was also used by two groups of students to create two applications in the context of a class. One of these applications automatically recreated derived files when input files changed, specifically recompiling an executable based on the input C files, inspired by past work on transparent make~\cite{transparentmake}. The other application performed automatic directory synchronization, syncing files in a local directory to remote storage. In these student projects, we found that Bento was robust enough to support a smooth development experience.


\begin{table*}[hbt!]
    \small
    \centering
    \begin{tabular}{l|>{\hspace{0pt}}r@{}r<{\hspace{0pt}}|>{\hspace{0pt}}r@{}r<{\hspace{0pt}}|>{\hspace{0pt}}r@{}r<{\hspace{0pt}}|r|>{\hspace{0pt}}r@{}r<{\hspace{0pt}}|r}
        \Xhline{4\arrayrulewidth}
        \textbf{Benchmark} & \multicolumn{2}{c|}{\normalsize
        \textbf{ext4-o}} & \multicolumn{2}{c|}{\normalsize
        \textbf{ext4-j}} & \multicolumn{2}{c|}{\normalsize
        \textbf{Bento-fs}} & \normalsize
        \textbf{Bento:ext4-j} & \multicolumn{2}{c|}{\normalsize
        \textbf{Bento-user}} & \normalsize
        \textbf{user:ext4-j} \\
        \hline
        \normalsize
        seq. read, 1-t, 4k & 286 & ($\pm$2) & 287 & ($\pm$2) & 289 & ($\pm$4) & \cellcolor{green!25}1.01 & 290 & ($\pm$2) & \cellcolor{green!25} 1.01 \\
        \normalsize
        seq. read, 1-t, 32k & 1811 & ($\pm$20) & 1796 & ($\pm$21) & 1817 & ($\pm$18) & \cellcolor{green!25} 1.01 & 1807 & ($\pm$18) & \cellcolor{green!25} 1.00 \\
        \normalsize
        seq. read, 1-t, 128k & 4170 & ($\pm$55) & 4071 & ($\pm$75) & 4119 & ($\pm$82) & \cellcolor{green!25} 1.01 & 4112 & ($\pm$50) & \cellcolor{green!25} 1.01 \\
        \normalsize
        seq. read, 1-t, 1024k & 6434 & ($\pm$129) & 6580 & ($\pm$197) & 6730 & ($\pm$197) & \cellcolor{green!25} 1.02 & 6510 & ($\pm$160) & \cellcolor{green!25} 0.99 \\
        \normalsize
        seq. read, 40-t, 4k & 429 & ($\pm$7) & 433 & ($\pm$9) & 436 & ($\pm$7) & \cellcolor{green!25} 1.00 & 429 & ($\pm$9) & \cellcolor{green!25} 0.99 \\
        \normalsize
        seq. read, 40-t, 32k & 3372 & ($\pm$65) & 3561 & ($\pm$332) & 3488 & ($\pm$184) & \cellcolor{green!25} 0.98 & 3417 & ($\pm$56) & \cellcolor{green!25} 0.96 \\
        \normalsize
        seq. read, 40-t, 128k & 17668 & ($\pm$143) & 17878 & ($\pm$162) & 17784 & ($\pm$132) & \cellcolor{green!25} 0.99 & 17833 & ($\pm$168) & \cellcolor{green!25} 1.00 \\
        \normalsize
        seq. read, 40-t, 1024k & 21407 & ($\pm$1774) & 22024 & ($\pm$101) & 22082 & ($\pm$339) & \cellcolor{green!25} 1.00 & 22136 & ($\pm$101) & \cellcolor{green!25}1.00 \\
        \normalsize
        rand. read, 1-t, 4k & 150 & ($\pm$1) & 149 & ($\pm$2) & 149 & ($\pm$2) & \cellcolor{green!25} 1.01 & 149 & ($\pm$3) & \cellcolor{green!25}1.00 \\
        \normalsize
        rand. read, 1-t, 32k & 1037 & ($\pm$6) & 1044 & ($\pm$6) & 1049 & ($\pm$8) & \cellcolor{green!25} 1.00 & 1041 & ($\pm$6) & \cellcolor{green!25} 0.99 \\
        \normalsize
        rand. read, 1-t, 128k & 2901 & ($\pm$20) & 2955 & ($\pm$36) & 2957 & ($\pm$33) & \cellcolor{green!25} 1.00 & 2908 & ($\pm$31) & \cellcolor{green!25}0.98 \\
        \normalsize
        rand. read, 1-t, 1024k & 5836 & ($\pm$68) & 5961 & ($\pm$152) & 5967 & ($\pm$116) & \cellcolor{green!25} 1.00 & 5890 & ($\pm$131) & \cellcolor{green!25}0.99 \\
        \normalsize
        rand. read, 40-t, 4k & 223 & ($\pm$24) & 211 & ($\pm$2) & 217 & ($\pm$5) & \cellcolor{green!25} 1.02 & 218 & ($\pm$5) & \cellcolor{green!25} 1.02 \\
        \normalsize
        rand. read, 40-t, 32k & 1717 & ($\pm$34) & 1712 & ($\pm$34) & 1737 & ($\pm$37) & \cellcolor{green!25} 1.01 & 1738 & ($\pm$31) & \cellcolor{green!25} 1.02 \\
        \normalsize
        rand. read, 40-t, 128k & 9265 & ($\pm$104) & 9232 & ($\pm$70) & 9206 & ($\pm$132) & \cellcolor{green!25} 1.00 & 9224 & ($\pm$55) & \cellcolor{green!25}1.00 \\
        \normalsize
        rand. read, 40-t, 1024k & 21635 & ($\pm$46) & 21650 & ($\pm$49) & 21637 & ($\pm$50) & \cellcolor{green!25} 1.00 & 21569 & ($\pm$54) & \cellcolor{green!25} 1.00 \\
        \hline
        \normalsize
        seq. write, 1-t, 4k & 234 & ($\pm$7) & 172 & ($\pm$3) & 252 & ($\pm$6) & \cellcolor{limegreen!50} 1.46 & 3.7 & ($\pm$0.0) & \cellcolor{orange!50} 0.02 \\
        \normalsize
        seq. write, 1-t, 32k & 860 & ($\pm$86) & 409 & ($\pm$1) & 1003 & ($\pm$65) & \cellcolor{limegreen!50} 2.45 & 4.0 & ($\pm$0.1) & \cellcolor{orange!50} 0.01 \\
        \normalsize
        seq. write, 1-t, 128k & 1058 & ($\pm$109) & 430 & ($\pm$44) & 1774 & ($\pm$352) &\cellcolor{limegreen!50} 4.12 & 4.0 & ($\pm$0.1) & \cellcolor{orange!50} 0.01 \\
        \normalsize
        seq. write, 1-t, 1024k & 1365 & ($\pm$0) & 469 & ($\pm$62) & 1843 & ($\pm$329) & \cellcolor{limegreen!50} 3.93 & 4.0 & ($\pm$0.0) & \cellcolor{orange!50} 0.01 \\
        \normalsize
        rand. write, 1-t, 4k & 142 & ($\pm$3) & 120 & ($\pm$1) & 139 & ($\pm$2) & \cellcolor{limegreen!50} 1.16 & 8.5 & ($\pm$0.14) & \cellcolor{orange!50} 0.07 \\
        \normalsize
        rand. write, 1-t, 32k & 875 & ($\pm$7) & 395 & ($\pm$22) & 898 & ($\pm$9) & \cellcolor{limegreen!50} 2.27 & 10.1 & ($\pm$0.0) & \cellcolor{orange!50} 0.03 \\
        \normalsize 
        rand. write, 1-t, 128k & 1952 & ($\pm$16) & 330 & ($\pm$18) & 2167 & ($\pm$62) & \cellcolor{limegreen!50} 6.55 & 10.3 & ($\pm$0.1) & \cellcolor{orange!50} 0.03 \\
        \normalsize 
        rand. write, 1-t, 1024k & 3051 & ($\pm$35) & 309 & ($\pm$8) & 3789 & ($\pm$56) & \cellcolor{limegreen!50} 12.24 & 10.1 & ($\pm$0.3) & \cellcolor{orange!50} 0.03 \\
        \normalsize 
        rand. write, 40-t, 4k & 230 & ($\pm$3) & 208 & ($\pm$4) & 241 & ($\pm$14) & \cellcolor{limegreen!50} 1.15 & 9.2 & ($\pm$0.1) & \cellcolor{orange!50} 0.04 \\
        \normalsize 
        rand. write, 40-t, 32k & 1237 & ($\pm$46) & 357 & ($\pm$61) & 1500 & ($\pm$34) & \cellcolor{limegreen!50} 4.20 & 10.0 & ($\pm$0.2) & \cellcolor{orange!50} 0.03 \\
        \normalsize 
        rand. write, 40-t, 128k & 1414 & ($\pm$43) & 303 & ($\pm$10) & 1894 & ($\pm$39) & \cellcolor{limegreen!50} 6.24 & 10.4 & ($\pm$0.1) & \cellcolor{orange!50} 0.03 \\
        \normalsize 
        rand. write, 40-t, 1024k & 1391 & ($\pm$49) & 296 & ($\pm$13) & 1924 & ($\pm$78) & \cellcolor{limegreen!50} 6.50 & 11.0 & ($\pm$0.0) & \cellcolor{orange!50} 0.04 \\
        \hline
        \normalsize 
        create, 1-t, ops/s & 12510 & ($\pm$418) & 8564 & ($\pm$186) & 12087 & ($\pm$390) & \cellcolor{limegreen!50} 1.41 & 194 & ($\pm$5) & \cellcolor{orange!50} 0.02 \\
        \normalsize 
        create, 40-t, ops/s & 34377 & ($\pm$2157) & 17858 & ($\pm$0) & 18819 & ($\pm$663) & \cellcolor{green!25} 1.05 & 216 & ($\pm$2) & \cellcolor{orange!50} 0.01 \\
        \hline
        \normalsize 
        delete, 1-t, ops/s & 23331 & ($\pm$878) & 22913 & ($\pm$0.3) & 24997 & ($\pm$0) & \cellcolor{green!25} 1.09 & 827 & ($\pm$11) & \cellcolor{orange!50} 0.03 \\
        \normalsize 
        delete, 40-t, ops/s & 60493 & ($\pm$7088) & 63253 & ($\pm$7101) & 57253 & ($\pm$6258) & \cellcolor{green!25} 0.91 & 808 & ($\pm$27) & \cellcolor{orange!50} 0.01 \\
        \Xhline{4\arrayrulewidth}
    \end{tabular}
    \caption{Performance results for ext4 in \texttt{data=ordered} mode (ext4-o), and \texttt{data=journal} mode (ext4-j), \system{}fs, and a userspace version of \system{}-fs (\system{}-user) on Filebench microbenchmarks using varying operation sizes and 1 and 40 threads.
    Reads and writes are measured in MBps. 
    Reads and writes are cached in the kernel and so can outperform the 2.5\,GBps and 2.0\,GBps device read and write speed.
    Results are averaged over 10 runs and standard deviations are included in parentheses.
    Color indicates performance relative to ext4-j. \system{}-fs performs similarly to ext4-j for most benchmarks. Both significantly outperform \system{}-user. 
}
    \label{tab:microbenchmark_performance}
\end{table*}


\section{Evaluation}


Our evaluation of \system{} aims to answer several questions: a) How well does \system{}-fs perform on different workloads? b) How robust is the file system under crash consistency testing? and c) How expensive are live upgrades?  

\subsection{Experimental setup}  

\noindent \textbf{Baselines.} 
We compare: a) ext4-o: ext4, the default file system on most Linux versions, using the default \texttt{data=ordered} option with metadata journaling, b) ext4-j: ext4 with data journaling (\texttt{data=journal} mode) c) \system{}-fs, and d) \system{}-fs running in userspace. We focus our evaluation on ext4 with journaling because \system{}-fs also implements data journaling. Note that \system{}-fs has implemented only a subset of ext4's optimizations. The userspace version of \system{} interacts with the storage device by opening it with the \texttt{O\_DIRECT} flag.


\noindent \textbf{Environment.}  
All experiments were run on a machine with Intel Xeon Gold 6138CPU (2 sockets, each with 20 cores, 40 hyperthreads), 96\,GB DDR4 RAM, and a 480\,GB Intel Optane SSD 900P Series with 2.5\,GB/s sequential read speed and 2\,GB/s sequential write speed. All benchmarks were run using the SSD as the backing device using the cores and memory on the socket connected to the SSD.

\subsection{Microbenchmarks}
\label{sec:microbenchmarks}
We ran microbenchmarks from the Filebench benchmarking suite. The workloads included sequential read, random read, sequential write, random write, and create and delete benchmarks. All workloads except for sequential write are run with both 1 thread and 40 threads. Read and write benchmarks were executed on a 4GB file using four different operation sizes: 4, 32, 128, and 1024KB. The create workloads create 800,000 16KB files in the same directory, allocating half before the start of the benchmark. The delete workloads delete 300,000 16KB files across many directories, with an average of 100 files per directory. All benchmarks were run 10 times, and averages and standard deviation were calculated.
\autoref{tab:microbenchmark_performance} shows the results on ext4 with both the default metadata journaling and data journaling, \system{}-fs, and \system{}-user, the userspace version of \system{}-fs. Results are colored based on the performance compared to ext4.

\textbf{Reads.} 
Reads on all three file systems have similar performance for all sizes and both single-threaded and 40-threaded, and large reads achieve greater bandwidth than provided by the device. This is because data is cached quickly after the first read, and all subsequent reads hit in the page cache. The userspace version uses the kernel cache in the FUSE kernel module before forwarding requests to userspace, so it performs similarly to direct kernel implementations. 

\textbf{Writes.} 
For small write benchmarks, \system{}-fs and ext4-j have fairly similar write performance. \system{}-fs has higher performance than ext4-j and similar performance to ext4-o on large write benchmarks due to slight implementation differences. Whereas ext4-j logs blocks to the journal on the write syscall path, \system{}-fs logs asynchronously in the writeback cache when data is flushed. This performance difference is more prominent for single-threaded benchmarks with large writes because these are more likely to stress the journal in ext4-j without stressing the writeback cache.  For all cases, the user-level implementation is much slower because it incurs additional kernel crossings and issues block I/O from userspace. Each operation must first pass from the kernel back to the userspace, which will then be translated into several read/write operations on the storage device.  Each system call to the device file must in turn pass through the VFS layer to reach the kernel block cache; this is much slower than direct accesses to the kernel block cache by a kernel file system. Additionally, \system{}-user does not have access to the JBD2 module, so it uses a simpler journal that is less efficient on large write workloads. This journal is also affected by slow userspace block I/O.


\textbf{Creates+Deletes.} 
On the create and delete benchmarks, ext4-j and \system{}-fs have similar performance. \system{}-fs outperforms ext4-j on single-threaded creates, likely due to the write speedup. Ext4-o outperforms \system{}-fs on multi-threaded creates. Both ext4 modes and \system{}-fs outperform the user-level file system for the same reason as the write benchmarks. 
\subsection{Application Workloads}

Next, we run three application-style workloads from Filebench, four applications, and two workloads each on two different key-value stores. All workloads were run 10 times and averages and standard deviation were calculated. 
From Filebench, we ran `varmail', `fileserver', and `webserver'.
(1) The `varmail' mail-serving workload uses 16 threads to create and delete 1000 files in one directory and performs reads and writes followed by fsyncs to these files.
(2) The `fileserver' file-serving workload uses 50 threads to create and delete 10,000 files across 500 directories and executes reads and appends to these files.
(3) The `webserver' web-serving workload uses 100 threads to read from 1000 small (16KB average size) files across around 50 directories and append to an operation log. All benchmarks execute for one minute.
For application workloads, we used `tar', `untar', and `grep' on the Linux kernel source code and `git clone' on the xv6 source repository.

We also evaluate read and write workloads on the Redis~\cite{redis} and RocksDB~\cite{rocksdb} key-value stores.
Redis is an in memory key-value store used in distributed environments. By default, it periodically dumps the database to a file but can be configured to also log all operations to an append-only-file (AOF) for persistence. In our evaluation, we use the AOF and configure it to sync every second. We run the `set' and `get' workloads from redis-benchmark, the provided benchmarking utility, for 1,000,000 operations using 100B values. RocksDB is a persistent key-value store developed by Facebook based on Google's LevelDB~\cite{leveldb}. Using db\_bench, the included benchmarking utility, we evaluate the `fillrandom' and `readrandom' workloads each for 1,000,000 operations using 100B values.

\begin{figure}
    \centering
    \includegraphics[trim=00 10 50 125, clip,width=0.38\textwidth]{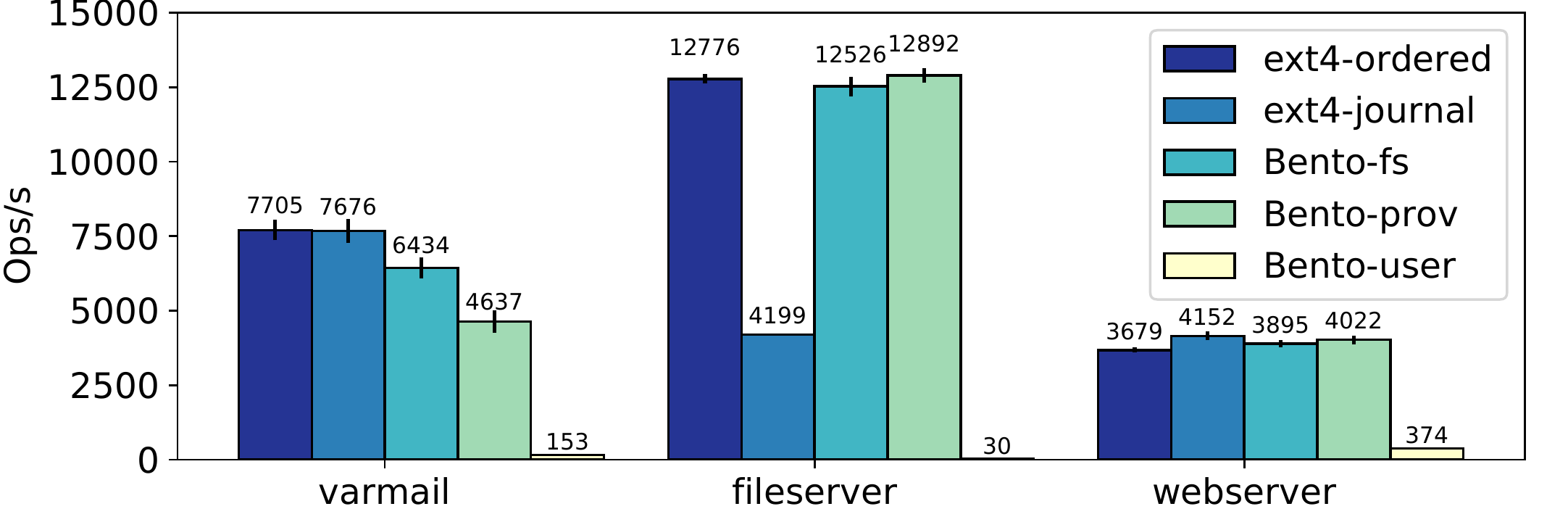}
    \caption{Performance results for ext4 in \texttt{data=ordered} mode and \texttt{data=journal} mode, \system{}-fs, \system{}-fs with provenance, and a userspace version of \system{}-fs on Filebench application-style workloads in ops/s. \system{}-user performs much worse on all benchmarks. \system{}-fs and \system{}-prov outperform ext4-journal on `fileserver' due to different handling of un-synced writes to deleted files.}
    \label{fig:macro}
\end{figure}

\textbf{Filebench: }\autoref{fig:macro} presents the application-style Filebench results for the three file systems described
earlier, plus \system{}-fs with file provenance (\system{}-prov).
Across all benchmarks, Bento-fs (with or without provenance) outperforms Bento-user by 10-400x
due to the reasons discussed earlier. 
For varmail and webserver, ext4-j and Bento-fs exhibit similar performance, but for fileserver, Bento-fs significantly outperforms ext4-j due to an unintentional quirk in the benchmark. 
Filebench `fileserver' executes many sequences of create-write-delete operations, but it does not sync the file before the file is deleted. With writeback caching, Bento recognizes that the pages
belong to files that no longer exist, and drops the writes. 
In ext4-j, on the other hand, writes are associated with the appropriate location on the storage device during the write syscall path by mapping the written page to the appropriate buffer head. This writeback code path therefore has no need to identify the written file and executes the block I/O regardless of whether the file exists or not. Like \system{}-fs, ext4-o is able to drop the writes to the deleted files so both file systems show similar performance.

\begin{figure}
    \centering
    \includegraphics[trim=00 20 60 225, clip,width=0.42\textwidth]{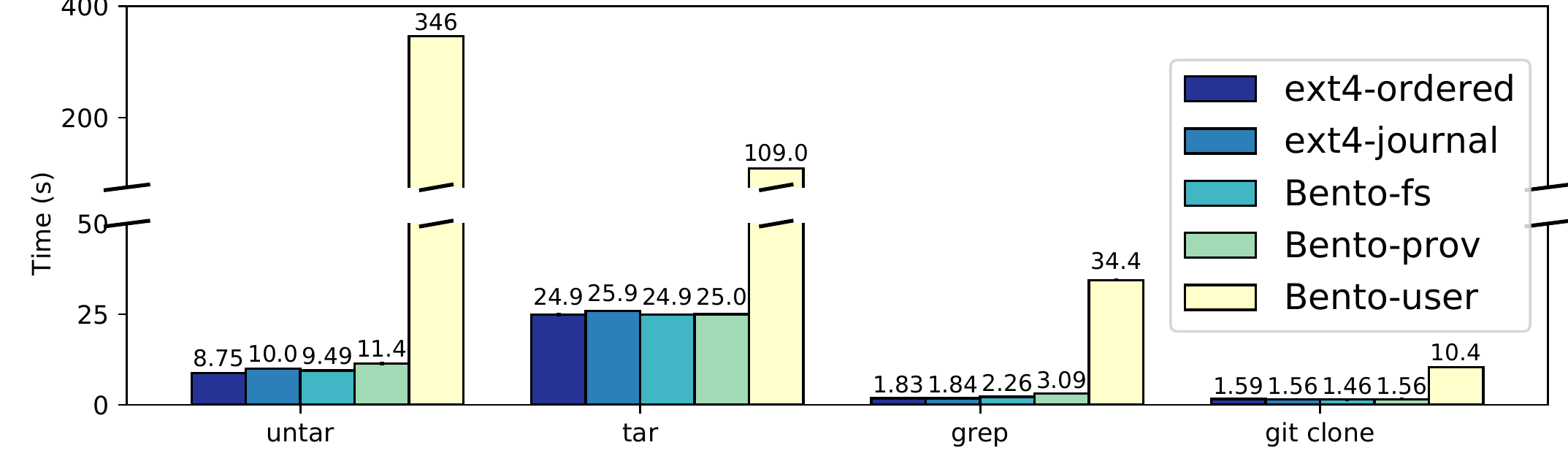}
    \caption{Performance results for ext4 in \texttt{data=ordered} mode and \texttt{data=journal} mode, \system{}-fs, \system{}-fs with provenance, and a userspace version of \system{}-fs on application workloads `tar', `untar', and `grep' on Linux source code and `git clone' on xv6. \system{}-user performs much worse than the other file systems. Ext4-journal performs somewhat better than \system{}-fs and \system{}-prov on `grep'.}
    \label{fig:application}
\end{figure}

\textbf{Applications: } \autoref{fig:application} shows the results for application workloads. Here, \system{}-fs outperforms \system{}-user by 4-36x. The difference is particularly noticeable for `untar' which involves many creates. Creates are particularly impacted by slow block I/O from userspace due to the large number of separate disk operations needed to modify the directory, allocate an inode, and fill the allocated inode. Relative to ext4-j, \system{}-fs performs similarly on `untar', `tar', and `git clone' and 19\% worse on  `grep'. The slowdown is due to optimized page caching in ext4 that is not implemented in \system{}-fs. Relative to ext4-o, \system{}-fs performs 13\% worse on `untar' due to data journaling and the lack of delayed allocation. On other benchmarks, ext4-o shows similar results to ext4-j.


For most tested workloads, Bento-prov has similar performance to Bento-fs. Bento-fs outperforms Bento-prov on `varmail' by 39\%, `untar' by 13\%, `grep' by 68\% because Bento-prov logs information on creates, deletes, opens, and closes. Similarly, Bento-prov is 25\% slower on the multithreaded create microbenchmark.

\begin{figure}
    \centering
    \includegraphics[trim=00 15 60 150, clip,width=0.42\textwidth]{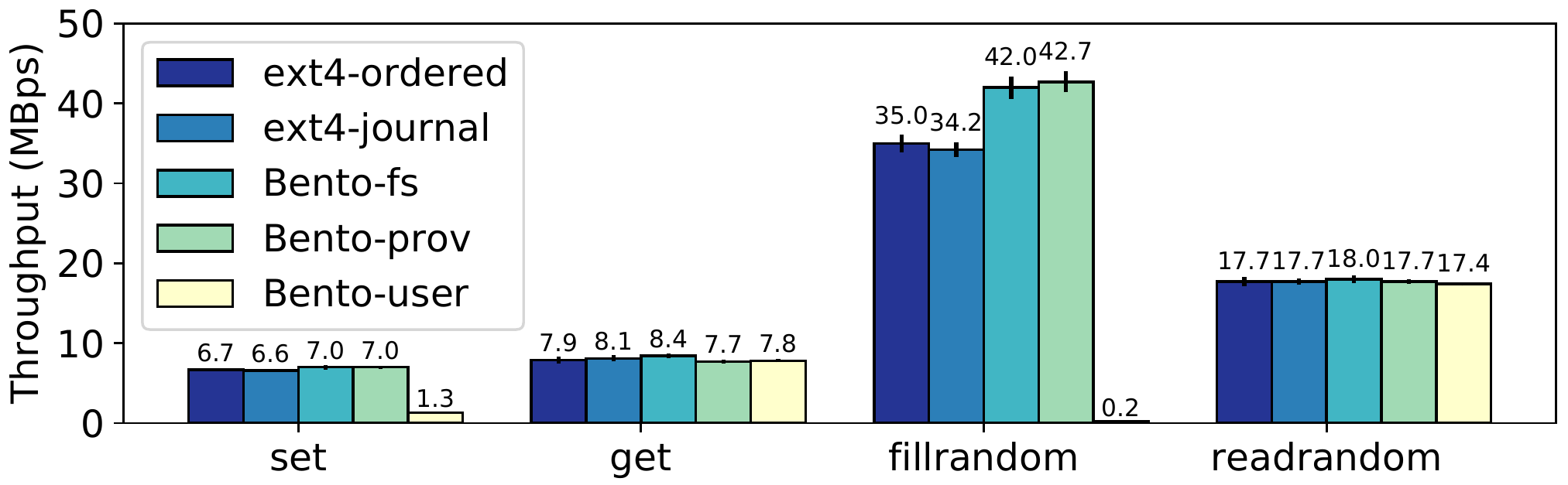}
    \caption{Performance results for ext4 in \texttt{data=ordered} mode and \texttt{data=journal} mode, \system{}-fs, \system{}-fs with provenance , and a userspace version of \system{}-fs on Redis `set' and `get' and RocksDB `fillrandom' and `readrandom'. \system{}-user performs much worse on write benchmarks.}
    \label{fig:kvstore}
\end{figure}

\textbf{Key-Value Stores: }
\autoref{fig:kvstore} shows
the results for Redis (`set' and `get') and RocksDB (`fillrandom' and `readrandom') workloads on the four file systems. 
Due to caching, \system{}-user performs similarly to the others on read-intensive workloads, 
but it performs much worse on writes. \system{}-fs and \system{}-prov show similar performance to ext4-j and ext4-o on reads but slightly outperform them on writes.

\begin{figure*}[t!]
\centering
    \begin{subfigure}[b]{0.38\textwidth}
    \centering
    \includegraphics[trim=0 0 00 180, clip,width=\textwidth]{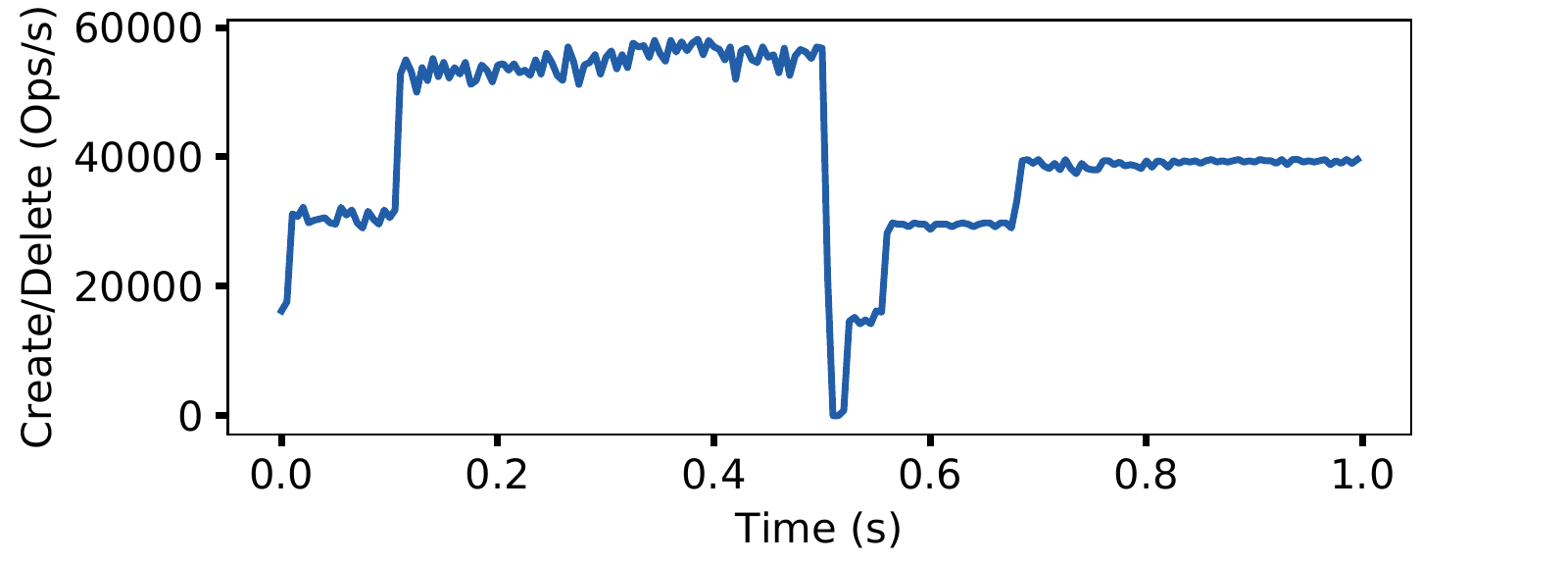}
    \caption{Create+delete operations in ops/s.}
    \label{fig:upgrade_create}
    \end{subfigure}
    \begin{subfigure}[b]{0.38\textwidth}
    \centering
    \includegraphics[trim=0 0 40 185, clip,width=\textwidth]{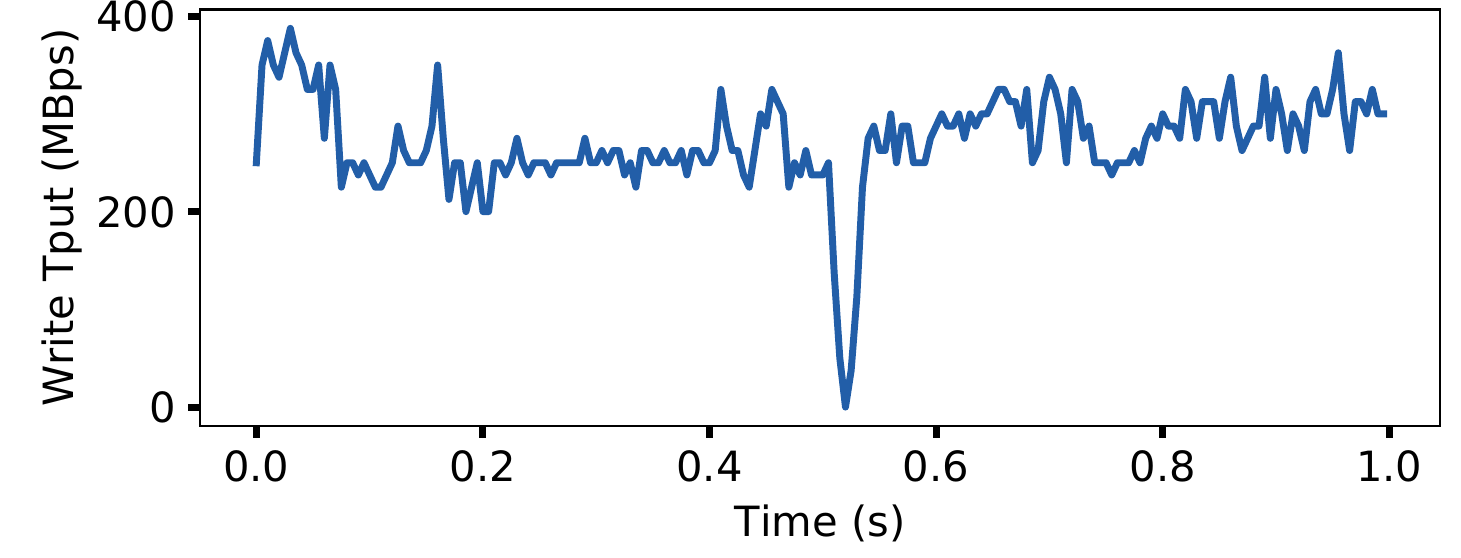}
    \caption{Synced writes with 10-threads in MBps.}
    \label{fig:upgrade_write}
    \end{subfigure}
    \caption{Performance during an upgrade from \system{}-fs to \system{}-prov, a provenance-tracking version of \system{}-fs. At 0.5 seconds, \system{}-fs is upgraded to \system{}-prov. The system experiences around 15ms of downtime.}
\end{figure*}

\subsection{Live Upgrade} 

In this section, we measure the effect of a live upgrade on application file system performance during an upgrade of the file system from Bento-fs to Bento-prov. We do not use Filebench for these benchmarks so we can collect latency of individual operations.
We ran two tests, both using a directory that initially contained 400,000 files. In the first, we executed a single thread that repeatedly created and deleted files. In the second, we executed 10 threads that repeatedly wrote and synced 64Kb writes to random files; we used 10 threads because with too many threads any service interruption caused by the upgrade was hidden by the latency variability of individual operations. In both tests, we upgraded to the version with provenance tracking after 0.5 seconds and completed the test after another 0.5 seconds. We converted the latency measurements into throughput by calculating the number of operations that occur each 5ms interval to smooth the data slightly. The results are shown in \autoref{fig:upgrade_create} and \autoref{fig:upgrade_write}.



These graphs show a performance drop where the upgrade occurred at 0.5 seconds. In both tests, the upgrade took around 15ms, during which time the file system was unavailable and a single operation per thread was blocked in the kernel. The performance recovered after the upgrade completed but create and delete performance was lower because the provenance-tracking file system performs extra work on these operations.

\section{Related Work}
\textbf{Using safe languages for kernel development.} 
Several systems, including Pilot~\cite{pilot}, SPIN~\cite{spin}, Singularity~\cite{singularity}, Biscuit~\cite{biscuit}, Redox~\cite{redox}, and Tock~\cite{tock} 
write the entire operating system, including the kernel, in a high-level language. 
SPIN leverages type safety to allow application-specific customization of kernel behavior.
We are also not the first to integrate Rust into the Linux kernel~\cite{rust_devices,linux_module_rust}.
The Berkeley Packet Filter (eBPF)~\cite{bpf} is a type safe language for safe extensibility in Linux. Users can insert eBPF programs at predefined kernel locations, and the kernel verifies the safety of the inserted programs before running them. 
ExtFUSE~\cite{extfuse} has enabled writing parts of a stackable file system using eBPF.
Compared to these, \system{} shows that it is possible to develop feature-rich file systems 
in a safe language to allow continuous integration of new features into a commodity operating system.

\textbf{Software fault isolation and verification.} An alternative approach is to allow 
development in an unsafe language (e.g., C) but do additional compile-time 
and runtime checks to prevent memory errors from affecting the rest of the system.
Software fault isolation (SFI)~\cite{sfi,bgi,lxfi} is a technique for sandboxing the impact of 
faults in C modules to the module itself; SFI has been widely used for protecting kernel device drivers.
We chose to use Rust instead as it has lower runtime overhead and provides the 
additional benefit of bug prevention in addition to sandboxing errors.
Software verification is a powerful tool for producing bug-free kernel code, and it has been shown that
a simple, single-threaded file system can be verified~\cite{yggdrasil}. Extending that work to
handle concurrency and high performance file systems is still ongoing.

\textbf{Moving kernel features to userspace.} Microkernel design, where kernel services run in userspace, is another way to speed operating system development~\cite{mach,l4} especially when safety
and/or development velocity are more important than raw performance.
Filesystem in Userspace (FUSE) is a good example in the Linux file system context. 
Many file systems have been developed in FUSE; when people need performance, they 
often re-implement the system inside the kernel~\cite{cephclient, glusterfs} using VFS.
With \system{}, developers no longer need to choose between performance and development velocity.

A related approach is to run the userspace OS service on dedicated processor cores, where 
applications communicate with the service via asynchronous message queues in shared memory~\cite{urpc,barrelfish,tas,snap}.
To date, this approach has only been proposed and not implemented for file systems~\cite{fsprocesses}. 
Performance can often be competitive with an equivalent kernel implementation, except when processors
need to busy wait or when the system needs page remapping for efficient zero copy I/O.

Rump kernels (or anykernels) enable running unmodified kernel code as userspace libraries by hijacking system calls and providing userspace implementations of necessary kernel internals. They are used for untrusted execution of kernel code, e.g., when mounting an untrusted file system, or userspace debugging.
Implementations exist for NetBSD as a rump kernel~\cite{RumpFS} and Linux as the  libOS~\cite{libos-linux} and Linux Kernel Library~\cite{lkl} projects; similarly,  User Mode Linux~\cite{uml} enables running a Linux kernel as a userspace process.

\textbf{OS live upgrade.}
There are three main commercially available tools for live upgrade of Linux systems:
 ksplice~\cite{ksplice-paper,ksplice}, kpatch~\cite{kpatch}, or kGraft~\cite{kgraft}. All three perform live upgrade of Linux kernel diffs and focus on security patches that do not modify data structure layout.
The internals of each approach differ, but all three reroute calls from modified functions to new functions.
Some research systems provide support for upgrade of more complex components.
Most similar to \system{}'s design is K42~\cite{k42}, a research operating system that enables upgrade of modular components by quiescing the component then transferring state to the new instance and updating references. 
PROTEOS~\cite{proteos}, another research operating system, also supports live upgrade of modular components.
DynAMOS~\cite{dynamos} and LUCOS~\cite{lucos} enable live upgrade of complex components in Linux without the need for state quiescence by using shadow data structures and virtualization, respectively, to maintain state.

\textbf{Stackable file systems.}
Stackable designs construct complex file systems by stacking layers of functionality on top of simple base file systems, enabling high velocity development.
File system stacking is natively supported by VFS and is used by the overlay file system and eCryptfs, but these file systems still suffer from the velocity problems caused by kernel C code.
FiST~\cite{fist} proposed a framework for development of portable stackable file systems written in a new high-level language, augmented with C code. This improves velocity by reducing the complexity of code written by developer, but cannot support complex file system data structures and cannot provide safety guarantees about the C code.

\section{Conclusion}

\system{} is a framework for high velocity development of Linux kernel file systems that enables several goals: safety, performance, generality, compatibility with existing operating systems, ability to do live upgrade, and support for easy debugging. 
\system{} provides these properties for file systems written in Rust, by translating Linux interfaces into safe interfaces with restricted memory sharing, 
supporting live upgrade with state transfer, and 
exposing identical interfaces to kernel and userspace file systems for userspace debugging.
We implement \system{}-fs, a simple file system using \system{} and show 
that it has similar performance to ext4 and significantly outperforms the version 
of \system{}-fs compiled to run in userspace. We develop a provenance tracking version of \system{}-fs,
and show that we can transparently upgrade \system-fs to it 
with only 15\,ms of service interruption to running applications.


\paragraph{Acknowledgements.}
We would like to thank Remzi Arpaci-Dusseau for his helpful feedback on earlier drafts of this paper.
We would also like to thank our anonymous reviewers and our shepherd, Rob Ross, for their helpful comments and feedback.
This work is partially supported by the National Science Foundation grant CNS-1856636 AM04. This work was also supported by Google and Huawei.

\bibliographystyle{plain}
\bibliography{main}

\clearpage

\thispagestyle{empty}
\begin{table*}[t]
    \centering
    \begin{tabular}{lp{60mm}}
        \Xhline{4\arrayrulewidth}
        \textbf{API Function} & \textbf{Description} \\
        \hline
        \textit{bento\_init(\&mut self, req, devname, fc\_info)} & Initialize the file system. \\
        \textit{bento\_destroy(\&mut self, req)} & Destroy the file system. \\
        \textit{bento\_lookup(\&self, req, parent, name, reply)} & Lookup a file \\
        \textit{bento\_forget(\&self, req, ino, nlookup)} & Forget lookups of a file \\
        \textit{bento\_getattr(\&self, req, ino, reply)} & Get attributes \\
        \textit{bento\_setattr(\&self, req, args..., reply)} & Set attributes \\
        \textit{bento\_readlink(\&self, req, ino, reply)} & Read a symbolic link \\
        \textit{bento\_mknod(\&self, req, parent, name, mode, rdev, reply)} & Create a file node \\
        \textit{bento\_mkdir(\&self, req, parent, name, mode, reply)} & Create a directory \\
        \textit{bento\_unlink(\&self, req, parent, name, reply)} & Unlink a file \\
        \textit{bento\_rmdir(\&self, req, parent, name, reply)} & Remove a directory \\
        \textit{bento\_symlink(\&self, req, parent, name, link, reply)} & Create a symbolic link \\
        \textit{bento\_rename(\&self, req, parent, name, newparent, newname, flags)} & Rename a file \\
        \textit{bento\_link(\&self, req, ino, newparent, newname, reply)} & Create a hard link \\
        \textit{bento\_open(\&self, req, ino, flags, reply)} & Open a file \\
        \textit{bento\_read(\&self, req, ino, fh, offset, size, reply)} & Read data from a file \\
        \textit{bento\_write(\&self, req, ino, fh, offset, data, flags, reply)} & Write data to a file \\
        \textit{bento\_flush(\&self, req, ino, fh, lock\_owner, reply)} & Called on each close of a file \\
        \textit{bento\_release(\&self, req, ino, fh, flags, lock\_owner, flush, reply)} & Called on the last close of an open file \\
        \textit{bento\_fsync(\&self, req, ino, fh, datasync, reply)} & Sync a file \\
        \textit{bento\_opendir(\&self, req, ino, flags, reply)} & Open a directory \\
        \textit{bento\_readdir(\&self, req, ino, fh, offset, reply)} & Read a directory \\
        \textit{bento\_releasedir(\&self, req, ino, fh, flags, reply)} & Called on the last close of a directory \\
        \textit{bento\_fsyncdir(\&self, req, ino, fh, datasync, reply)} & Sync a directory \\
        \textit{bento\_statfs(\&self, req, ino, reply)} & Get file system statistics \\
        \textit{bento\_setxattr(\&self, req, ino, name, value, flags, position, reply)} & Set extended attributes of a file \\
        \textit{bento\_getxattr(\&self, req, ino, name, size, reply)} & Get extended attributes of a file \\
        \textit{bento\_listxattr(\&self, req, ino, size, reply)} & List extended attributes of a file \\
        \textit{bento\_removexattr(\&self, req, ino, name, reply)} & Remove an extended attribute of a file \\
        \textit{bento\_access(\&self, req, ino, mask, reply)} & Check file permissions \\
        \textit{bento\_create(\&self, req, parent, name, mode, flags, reply)} & Create and open a file \\
        \textit{bento\_getlk(\&self, req, ino, fh, lock\_owner, start, end, typ, pid, reply)} & Test for a file lock \\
        \textit{bento\_setlk(\&self, req, ino, fh, lock\_owner, start, end, typ, pid, sleep, reply)} & Acquire a file lock \\
        \textit{bento\_bmap(\&self, req, ino, blocksize, idx, reply)} & Map a block index within a file \\
        \textit{bento\_update\_prepare(\&mut self) -> Option<TransferOut>} & Prepare to be removed during a live upgrade \\
        \textit{bento\_update\_transfer(\&mut, Option<TransferIn>)} & Initialize during a live upgrade \\
        \Xhline{4\arrayrulewidth}
    \end{tabular}
    \caption{The full File Operations API, based on the FUSE lowlevel API with \textit{bento\_update\_prepare} and \textit{bento\_update\_transfer} added for live upgrade. File systems implement a subset of the provided functions. The \textit{req} includes the requesting application's user id, group id, and process id. The \textit{reply} data structures are used to return data or error values.} 
    \label{tab:fileops-api-long}
\end{table*}


\end{document}